\documentclass[twocolumn]{aastex631}

\usepackage{footmisc}
\usepackage{tablefootnote}
\usepackage{xspace}
\usepackage[utf8]{inputenc}
\usepackage[T1]{fontenc}
\usepackage{natbib}

\usepackage{amsmath}

\long\def\ltx@foottext#1#2{%
\begingroup 
\expandafter\ltx@make@current@footnote\expandafter{\@mpfn}{#1}%
\@footnotetext{\vtop{\iftwocolstyle\hsize=.5\textwidth 
\advance\hsize-18pt 
\fi #2\vskip2pt}}
 \endgroup 
}

\def\vplanet{\texttt{\footnotesize{VPLanet}}\xspace}
\def\atmesc{\texttt{\footnotesize{AtmEsc}}\xspace}
\def\flare{\texttt{\footnotesize{FLARE}}\xspace}
\def\stellar{\texttt{\footnotesize{STELLAR}}\xspace}

\begin{document}

\title{The Impact of Stellar Flares on the Atmospheric Escape of Exoplanets orbiting M stars I: Insights from the AU Mic System.}

\correspondingauthor{Laura N. R. do Amaral}
\email{laura.nevesdoamaral@gmail.com, lamaral1@asu.edu}

\author[0000-0002-0786-7307]{Laura N. R. do Amaral}
\affiliation{School of Earth and Space Exploration, Arizona State University \\
781 Terrace Mall, Tempe, AZ 85287, USA}
\affiliation{Consortium on Habitability and Atmospheres of M-dwarf Planets (CHAMPs), Laurel, MD, USA}
\affiliation{NASA Virtual Planetary Laboratory Lead Team, USA}

\author[0000-0002-7260-5821]{Evgenya L. Shkolnik}
\affiliation{School of Earth and Space Exploration, Arizona State University \\
781 Terrace Mall, Tempe, AZ 85287, USA}
\affiliation{Consortium on Habitability and Atmospheres of M-dwarf Planets (CHAMPs), Laurel, MD, USA}
\affiliation{NASA Virtual Planetary Laboratory Lead Team, USA}

\author[0000-0001-5646-6668]{R. O. Parke Loyd}
\affiliation{Eureka Scientific, 2452 Delmer Street Suite 100, Oakland, CA, 94602-3017, USA}

\author[0000-0002-1046-025X]{Sarah Peacock}
\affiliation{Consortium on Habitability and Atmospheres of M-dwarf Planets (CHAMPs), Laurel, MD, USA}
\affiliation{University of Maryland, Baltimore County, Baltimore, MD, 21250, USA}
\affiliation{NASA Goddard Space Flight Center, Greenbelt, MD 20771, USA}

\shorttitle{Effect of Flares on Primordial Atmospheres.}

\begin{abstract}

The X-rays and Extreme Ultraviolet (XUV) emission from M stars can drive the atmospheric escape on planets orbiting them. M stars are also known for their frequent emission of stellar flares, which will increase the high-energy flux received by their orbiting planets. To understand how stellar flares impact the primordial atmospheres of planets orbiting young M stars, we use UV spectroscopic data of flares from the Habitable Zones and M dwarf Activity across Time (HAZMAT) and Measurements of the Ultraviolet Spectral Characteristics of Low-mass Exoplanetary Systems (MUSCLES) programs as a proxy to the XUV flare emission. Using the software package \vplanet, we simulate the young AU Mic planetary system composed of two Neptune-sized and one Earth-sized planet orbiting a 23-Myr-old M1 star. Our findings show that the Earth-sized planet AU Mic d should be in the process of losing completely its atmosphere in the next couple million years, solely due to the quiescent emission, with flares not significantly contributing to its atmospheric escape due to the small size of AU mic d and its close-in distance from the star. However, our results indicate that flares would play a crucial role for such planets further away, in the habitable zone (i.e. 0.2935 AU) of AU Mic-like stars during the post-saturation phase, accelerating the total atmospheric loss process by a few billion years. For planets between 0.365 AU and the HZ outer edge, the additional XUV from flares is necessary to deplete primordial atmospheres fully since the quiescent emission alone is insufficient.

\end{abstract}


\section{Introduction} \label{sec:intro}

The high-energy radiation from a star, in the range of Extreme Ultraviolet (EUV) and X-rays, (XUV; 0.5-117 nm, for the purposes of this work), is known to affect exoplanetary atmospheres by changing their chemistry and driving escape \citep{murray2009atmospheric,des2010evaporation,luger2015habitable,zhang2022detection}. However, on top of the high-energy quiescent emission, stellar activity in the form of flares increases the amount of radiation that stars can release and deliver to planetary atmospheres \citep{hawley2014kepler,yamashiki2019impact,johnstone2020hydrodynamic,hu2022extreme,feinstein2024hst}. Previous work has shown that stellar flares can enhance atmospheric erosion \citep{chadney2017effect,lee2018effects,france2020high,atri2021stellar} and water loss \citep{amaral2022} in non-magnetized planets. Stellar flares might also promote photochemical changes in the atmospheres of planets \citep{venot2016influence,tilley2019modeling,chen2021persistence}, including the depletion of the ozone layer of terrestrial planets \citep{segura2010effect}.

Observations have shown that stellar flare rate decreases with age \citep{davenport2019evolution,loyd2021hazmat,engle2023living,2024Feinstein}, which leads to exoplanets experiencing a change in their space weather conditions over time. M stars in particular exhibit higher flare rates compared to FGK stars \citep{shibayama2013superflares, hawley2014kepler,gunther2020stellar} across all ages. The high-energy environment of M stars, aligned with the fact that these stars host 27 of the 29 rocky planets that are considered most similar to the Earth (according to PHL @ UPR Arecibo; see \citealt{schulze2011two} for more), makes them important targets for which to explore the influence of space weather in planetary atmospheres. 

To better understand the space weather environment around active M stars, we need to know how much energy stellar flares emit. Stellar flares vary by orders of magnitude in their total energy. Because they are randomly timed and of variable strength, they are best described through statistical distributions. Flare Frequency Distributions (FFDs) represent the rate of flares as a function of the energy they release (or another measure of amplitude). Higher energetic events occur less frequently than lower-energy events, and it is common to adopt a power law to approximate the FFD of a star. The parameters of these power-law FFDs (slope and Y-intercept) depend not just on the star in question but on the wavelength band of interest. Therefore, simulating atmospheric loss due to flares requires FFDs specific to flare emission in the XUV wavelength range.

Modeling the FFD evolution over time can be complex, however, especially for M stars. The nuclear reactions in the cores of these stars occur at a slow rate due to their low mass (0.08 - 0.6 M$_{\odot}$). This characteristic causes their physical properties to change very slowly, making age determination of isolated M stars a challenge (see \citealt{soderblom2010ages} for more references). Despite this, previous work has explored the relationship between stellar age and flare activity on M stars. \citet{davenport2019evolution} presented a stellar mass-age-dependent FFD model for FGKM stars from the Kepler catalog \citep{davenport2016kepler} showing that flare activity does not change considerably over time. Using TESS observations and ages calculated from cluster association, \citet{ilin2021flares} showed that those results overestimate the FFD for M stars by two orders of magnitude. The most probable source of error is related to age constraints and the low number of M stars in the Kepler sample when compared to the TESS targets. Also using optical data of TESS, \citet{feinstein2024hst} shows that the FFD slope of M stars is mainly constant over the first 200 Myrs.

Another challenging factor is how to build age-dependent FFD models in the XUV range, where observations are scarce. Even with the abundant detection of flares by TESS and Kepler \citep{davenport2016kepler,gunther2020stellar,2020Feinstein,pietras2022statistical,feinstein2024hst}, these observations were made in the optical range. \citet{Jackman2024} explored the relation between Far-Ultraviolet (117 - 143 nm, FUV; \citealt{2018Loyd}) and optical flares, showing that the results from \citet{davenport2019evolution} underestimate the evolution of the FUV emission due to flaring and that optically quiet M stars still present flares in the FUV. However, \citet{macgregor2021discovery} showed that the same flare in the M5 V star Proxima Centauri in the FUV with the Hubble Space Telescope (HST) presented a flux 1500 times higher than the flux observed in the optical by TESS. These results show evidence that stellar flares cannot be scaled directly from the optical range to the FUV or more energetic wavelengths, such as XUV \citep{jackman2023extending,brasseur2023constraints}.

Unlike the optical range, FUV and XUV radiation can cause molecular dissociation, alter the photochemistry of planetary atmospheres, and drive atmospheric escape, which affect planetary evolution, habitability, and characterization. These points emphasize the need for flare observations in the XUV, which is a current limitation in creating an FFD model that does not rely on multiple flare energy scaling factors or scaling factors based on observations in the optical. To minimize uncertainties, we use FFDs measured in the FUV as a proxy of XUV. Whereas the emission from stellar flares in the optical range comes primarily from the photosphere, the high-energy wavelengths are emitted primarily from the upper atmosphere (chromosphere, transition region, and corona), so FUV and XUV are emitted through the same physical process, and thus they are intrinsically tied \citep{1997Dere,2024Dufresne}.

Building a reliable XUV FFD for stars of different ages is, however, currently impossible due to the lack of observations in the EUV. Less than 10 M stars have flares observed in the EUV range \citep{2000Audard,2003Gudel,2003Hawley,2022Feinstein}.
Besides that, the hydrogen present in the interstellar medium absorbs much of the EUV  \citep{craig1997extreme,redfield2008structure,wood2005stellar}, making it difficult to observe it for nearby stars and impossible for distant stars. 

Considering all these aspects, we interpolate the observed FFDs from the \textit{HAbitable Zones and M dwarf Activity across Time} (HAZMAT; \citealt{2018hazmativ})  and Measurements of the Ultraviolet Spectral Characteristics of Low-mass Exoplanetary Systems (MUSCLES) \citep{2016LoydMUSCLES,france2016muscles,2016YoungbloodMUSCLES,2018Loyd} projects over time to build a more accurate age-dependent FFD model (see Section \ref{subsec:ffd}). HAZMAT is a program focused on observing the evolution of the quiet and flaring M and K stars in X-rays, FUV, and Near-Ultraviolet (NUV), providing spectral data of these stars using observations from ROSAT, GALEX, HST, and semi-empirical, full-wavelength models. The data used to build the FFDs by \citet{2018hazmativ} are in the FUV range facilitating the scaling to the XUV range, in contrast with FFDs based on optical observations (e.g., TESS). MUSCLES is a panchromatic spectral survey (from infrared to X-ray) focused on spectral data for M and K stars that host exoplanets, using Chandra, X-ray Multi-Mirror-Newton (XMM-Newton), Swift, HST, and Differential Emission Measure models (DEM).

To understand how the XUV emission from flares impacts the planets close-in to M stars, we simulated the primordial hydrogen atmospheric evolution of the planets AU Mic b, c, and d \citep{plavchan2020planet,2021Martioli,2023WittrockAUMic}. Due to its young age of 23 $\pm$ 3 Myrs \citep{mamajek2014age,2017Shkolnik}, the AU Mic system presents a unique opportunity to understand the first stages of a planetary system's evolution. Also, the presence of a likely Earth-size planet, AU Mic d, with a mass of 1.053 $\pm$ 0.511 M$_{\oplus}$ \citep{2023WittrockAUMic} opens the possibility of exploring the system's habitability. AU Mic is an active M1 star \citep{2021Klein} with detected flares \citep{1993Robinson,1994Cully,2022Feinstein} that can reach about 10$^{36}$ ergs in the EUV \citep[7-38 nm]{1994Cully}, and a cumulative FFD of $log \ \nu = 23.42 (\pm4.78) \ logE_{U} \ - \ 0.72(\pm 0.15)$ in the NUV (303–417 nm; \citet{tristan20237}). The AU Mic's planets are located between the star and the inner edge of the conservative habitable zone (HZ, \citet{2014Kopparapu}), that is, out of the HZ. Because of this, these planets receive more XUV radiation than if they were within the HZ limits, which for M stars is $\geq$ 1.85 W/m$^{2}$ \citep{richey2023hazmat}, equivalent to 400x the XUV flux presently received by the Earth (0.00464  W/m$^{2}$, \citet{luger2015habitable}). Lyman-$\alpha$ transits have shown that AU Mic b, a Neptune-like planet with a planetary mass and radii of 10.2$_{-2.7}^{+3.9}$ M$_{\oplus}$ and 4.07$\pm$0.17 R$_{\oplus}$ \citep{2023Donati,2021Martioli}, is currently undergoing atmospheric loss of neutral hydrogen, with an estimated loss rate of 2.25–4.44 $\times$ 10$^{11}$ g/s \citep{rockcliffe2023variable}. However, it is not clear if the observed phenomenon is due to interaction with the constant wind from the star or is caused by XUV photoionization, which gives us the opportunity to explore the possible scenarios using computational modeling. In Section \ref{sec:methodology}, we describe the atmospheric escape model and how we built an FFD age-dependent model, using the FFDs from \citet{2018hazmativ}. In Section \ref{sec:results}, we show the results for the atmospheric evolution of AU Mic b, c, and d and how the flaring evolution from AU Mic impacts them. Finally, in Section \ref{sec:conclu}, we discuss and summarize our findings.
\section{Methodology} \label{sec:methodology}

Our simulations consider three principal components: the stellar evolution of AU Mic, the evolution of its FFD, and the resulting planetary atmospheric escape of a primordial hydrogen envelope. To perform these simulations, we use the code \vplanet\footnote{\url{https://github.com/VirtualPlanetaryLaboratory/vplanet}} \citep{barnes2020vplanet}. \vplanet is a modular code, where each module is responsible for different aspects of planetary habitability, such as orbital dynamics, planetary interior, stellar evolution, and atmospheric escape. This modular functionality in the code allows us to perform simulations combining two or more features and coevolving parameters (i.e., the luminosity calculated by the stellar evolution model is converted to stellar flux when it reaches the planet and is used to calculate the planetary atmospheric escape). \vplanet also has an adaptative time step, which helps to optimize the simulation time (see Table \ref{tab:simulations} for the range of time steps allowed in the simulations of this work).

\subsection{Stellar Evolution}

\begin{figure}
\centering
\includegraphics[scale=0.2]{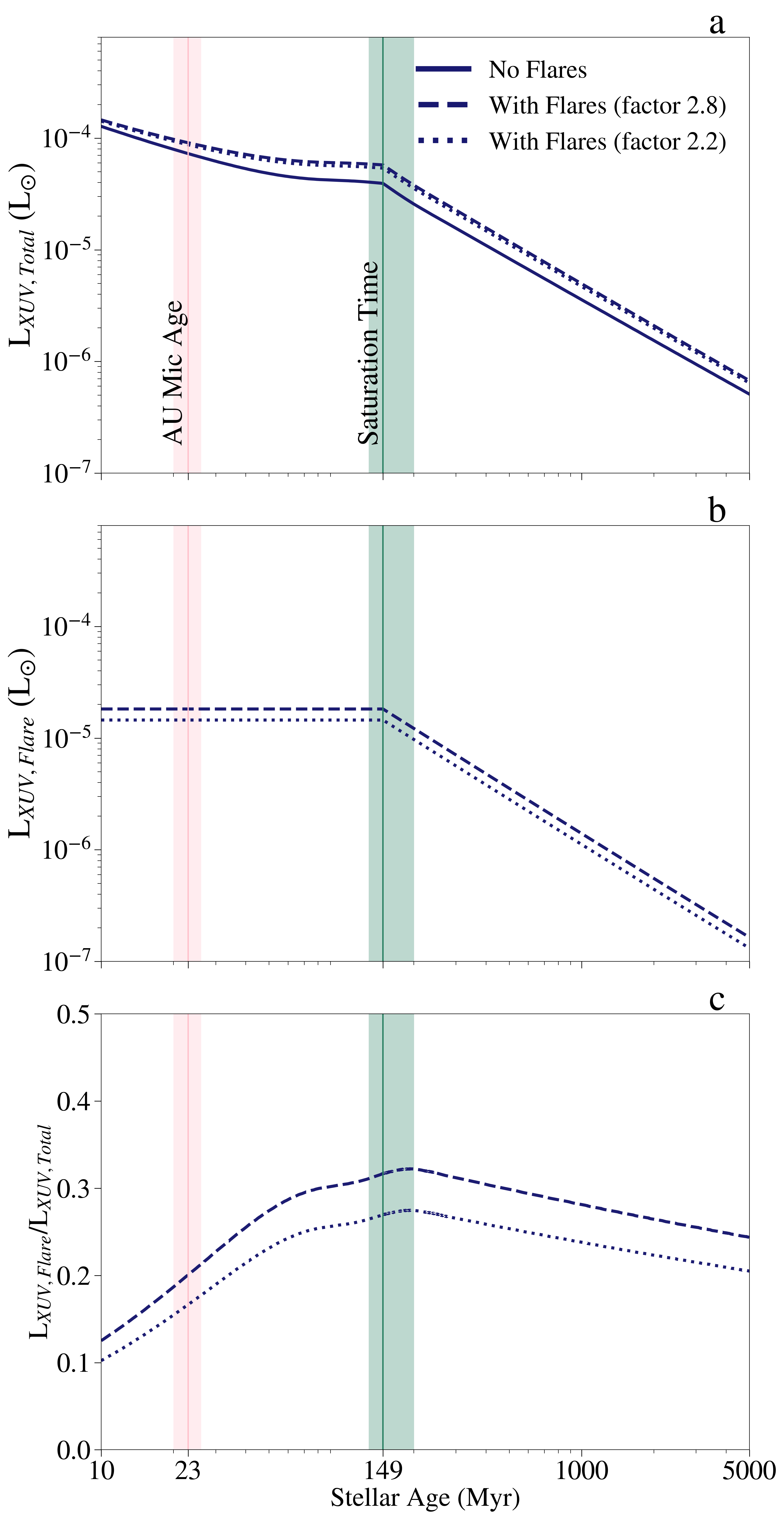}
\caption{Stellar XUV evolution of AU Mic. Panel a is the total XUV luminosity, where dashed and dotted lines are the sum of quiescent and flaring XUV, and the solid line is the quiescent alone. Panel b shows the flaring XUV luminosity evolution, and panel c shows the ratio between the flaring XUV luminosity and the total XUV luminosity (i.e., quiescent plus flaring). The vertical green line represents the end of the star's saturation phase, and the pink represents the current age of AU Mic, both with their associated uncertainties shaded. The results using \citet{2018Loyd} (i.e. using scaling factor from FUV to XUV equal to 2.8) and \citet{2022Feinstein} (i.e. using scaling factor from FUV to XUV equal to 2.2) the quiescent XUV evolution from \citet{2005Ribas} are represented by dashed, dotted, and solid lines, respectively.}
\label{fig:star}
\end{figure}

To perform the simulations, we use the \stellar module from \vplanet, which interpolates the stellar evolutionary models from \citet{baraffe2015new}. That module output provides the time evolution of the stellar bolometric luminosity $L_{bol}$ (Fig. \ref{fig:star}.a), stellar effective temperature $T_{eff}$, and stellar radius $R_{*}$. The quiescent XUV emission follows the \citet{2005Ribas} model (Fig. \ref{fig:star}.b), and it is given by

\begin{equation}\label{eq:LXUV}
    \frac{L_{XUV}}{L_{bol}} =  \begin{cases} f_{sat} &  t \leq t_{sat} \\
                                 f_{sat}\Big(\frac{t}{t_{sat}}\Big)^{-\tau_{XUV}}  & t > t_{sat}\\      
                                  \end{cases}
\end{equation}

The ratio between the quiescent XUV emission $L_{XUV}$ and the bolometric luminosity $L_{bol}$ will remain constant ($f_{sat}$) until the end of the saturation phase (i.e., the star age $t$ reaches the saturation time $t_{sat}$, $t = t_{sat}$). After the end of the saturation phase (i.e., during the post-saturation regime), the stellar activity due to upper atmospheric XUV emission starts to decline inversely proportional to the $\tau_{XUV}$ coefficient (second line of Equation \ref{eq:LXUV}).

Observations have shown that the duration of the saturation phase in early M stars is on the order of a few hundred million years \citep{stelzer2013uv,shkolnik2014hazmat,richey2023hazmat}. A star such as AU Mic has a saturation phase of approximately 149$^{+51}_{-19}$ Myr \citep{richey2023hazmat}. 

\subsection{Age-Dependent Flare Frequency Distribution}\label{subsec:ffd}
\begin{figure}
\centering
\includegraphics[scale=0.44]{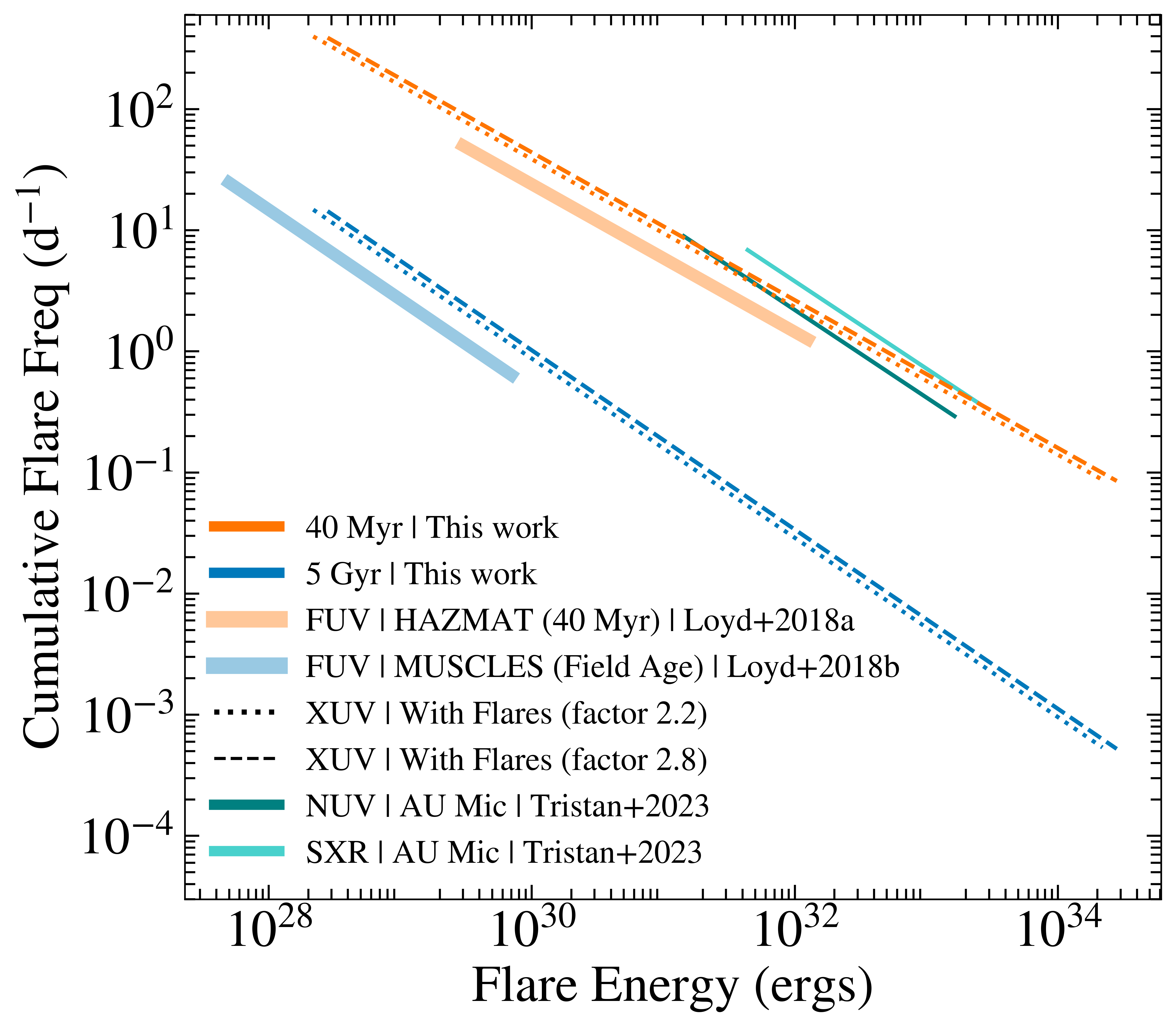}
\caption{FFD used in this work. The FFD in the FUV from the HAZMAT and MUSCLES \citep{2018hazmativ,2018Loyd} data (solid lines with low opacity) was scaled to XUV (during flaring) using the MUSCLES conversion factor (i.e., 2.8, see Table \ref{tab:convert}; dashed lines) and to the XUV (in the quiet state) using the stellar spectra of AU Mic from \citet{2022Feinstein} (i.e., 2.2, see Table \ref{tab:convert}; dotted lines). The Blue color represents stars with field age ($\sim$ 5 Gyrs), and the orange represents young stars (40 Myrs). The teal and turquoise lines are the observed AU Mic FFD \citep{tristan20237} in the NUV (303-417 nm) and Soft X-rays (0.103-3.5 nm), respectively.}
\label{fig:ffd}
\end{figure}

The flaring observations in the FUV range of the HAZMAT and MUSCLES project were done using spectroscopic data from the HST \citep{2018hazmativ,2018Loyd}. Their observations produced FFDs in the FUV for stars at the age of 40 Myr (HAZMAT) and 1-5 Gyrs (MUSCLES), represented by low opacity solid lines in Fig. \ref{fig:ffd}. We interpolate these FFDs over stellar age and build an age-dependent FFD model. The HAZMAT FFD is based on flares averaged across the observations of 12 stars spanning effective temperatures of 3506-3990 K, and the MUSCLES FFD is based on 10 stars spanning 3098-3679 K, encompassing the effective temperature of AU Mic, 3665 $\pm$ 31 K found by \citet{2023Donati}. We perform a linear interpolation between the two equations from \citep[Fig. 1]{2018hazmativ}, considering the age of the young stars as 40 Myrs and the field age as 5 Gyrs. Here, we are considering that the stellar activity does not change significantly between 23 Myrs (AU Mic’s age) and 40 Myrs so that we can consider the FFD at 40 Myrs as the same as 23 Myrs. We make this assumption because the flare activity does not present substantial changes during the saturated phase \citep{2024Feinstein}. Similar to \citet{2018hazmativ}, we define the flare rate in the FUV as $\nu_{FUV}$ in flares per day, and the energy $E_{FUV}$ in 10$^{30}$ ergs. The fit provides the FFD slopes in function of stellar age, $t$ (in Myrs), and is given by

\begin{align}\label{eq:ffd}
    log (\nu_{FUV}) = \alpha \ log (E_{FUV}) + \beta_{FUV}
\end{align}

Where $\alpha$ and $\beta_{FUV}$ are given by 

\begin{equation}
    \alpha=  \begin{cases} -0.61 &  t \leq t_{sat} \\  
                                  -0.085\ log10(t/t_{sat})-0.61 & t > t_{sat}\\
                                  
                                  \end{cases}\\
\end{equation}
\begin{equation}
    \beta_{FUV}=  \begin{cases} 1.38 &  t \leq t_{sat} \\ 
                                  -1.108\ log10(t/t_{sat})+1.38 & t > t_{sat}\\
                                
                                  \end{cases}
\end{equation}

The flare contribution is calculated using Equation \ref{eq:ffd} to estimate the XUV luminosity from flares $L_{XUV,f}$. We convert the flare rates from FUV to XUV (see Section \ref{sec:scaling}) and then calculate the XUV luminosity from flares as

\begin{equation}
       L_{XUV,f} = \int_{E_{min}}^{E_{max}} \nu(E_{XUV,f})E_{XUV,f}\ dE,
\label{eq:LXUVFlare}
\end{equation}
where $E_{\mathrm{XUV},f}$ is the XUV energy of the flare, and $\nu(E_{XUV,f})$ is the flare rate (number of flares per unit of time) for each flare with energy $E_{XUV,f}$. $E_\mathrm{min}$ and $E_\mathrm{max}$ as the minimum and maximum flare energy considered in the calculations. The values considered in the simulations span from 10$^{28}$ to 10$^{34}$ ergs (see Table \ref{tab:simulations}), following the observed flare energy values from AU Mic \citep{2018Loyd,tristan20237}. Flare observations from \citet{tristan20237} show that the upper limit for the flare energies on AU Mic in the NUV range is about 1.6 $\times$ 10$^{33}$ ergs, which we can translate to 2.4 $\times$ 10$^{33}$ ergs in the FUV, based on the scaling factor found by \citet{jackman2023extending}. For the purpose of this work, we round this value to 10$^{34}$ ergs and set it as the upper limit for the flare energy in the simulations. From Equation \ref{eq:LXUVFlare}, we calculate the flux reaching the planets in each time step. The time step in the simulations varies from days to years automatically. This means that we do not simulate a series of flares, but instead we calculate the average flux from flares in each time step.

Since our goal is to model the evolution of a planetary system over billions of years, the statistical data over time that we use to model the age-dependent FFD allow us to average out the flux of flares in the simulations. By averaging out the flux emitted by flares over time, we are also keeping the system in an efficient atmospheric escape regime (i.e., energy-limited regime, see Section \ref{sec:atmesc}), which leads to a higher amount of atmosphere lost at the end of the simulations when compared to the real case where the stellar flares reach the atmosphere one-by-one. This assumption implies that our results reflect an upper limit of the atmospheric loss experienced by the AU Mic planets. To resolve single flare events in our simulations, we would need to consider a time step on the order of minutes, which would require a long and impractical computational time. Also, our simulations aim to understand the long-term impact of stellar flares on primordial planetary atmospheres over billions of years, dispensing the necessity of a time resolution lower than days.

\subsection{Scaling factors}\label{sec:scaling}

Flares in the FUV band used to build the HAZMAT and MUSCLES FFDs promote atmospheric heating but are not capable of ionizing hydrogen atoms due to the low energy, which is below the energy threshold of hydrogen ionization ($>$ 13.6 eV). As presented in Section \ref{sec:atmesc}, we calculated the atmospheric escape using two approaches, the energy-limited and radiation-recombination limited regimes of XUV-powered hydrodynamic escape. To find the corresponding values of XUV flux due to flaring reaching the AU Mic's planets, we use two scaling factors to convert the FUV FFDs to the XUV bandpass (see Table \ref{tab:convert}). 

The first scaling relationship is based on the energy budget of the fiducial flare\footnote{\url{https://github.com/parkus/fiducial_flare}} model developed through observations made by the MUSCLES program \citep{2018Loyd}. The fiducial flare energy budget is based on the averaged flares of 10 M stars observed by the HST from 117 to 143 nm, Far-Ultraviolet Spectroscopic Explorer (FUSE) observations of AD Leo from 92 to 117 nm, and Extreme ultraviolet Variability Experiment (onboard the Solar Dynamics Observatory) observations of solar flares from 10 to 92 nm, scaled according to appropriate line ratios \citep{2018Loyd}. We integrated this energy budget from 10 to 117 nm and took the ratio between the flux in this wavelength range and the flux integrated across the FUV band defined above to obtain an XUV/FUV ratio (scaling factor) of 2.8. Although the fiducial flare energy budget does not extend all the way to the 0.5 nm wavelength that defines the lower limit of the XUV range used by VPlanet, we expect the additional energy in this range to be minor. For comparison, this range accounts for 15$\%$ of the total energy in the 0.6-117 nm range in the MUSCLES spectrum of the M1V star GJ 832, which includes X-ray observations and reconstructed X-ray spectrum covering the 0.6-10 nm range.

The second scaling factor was estimated using the stellar spectral data from AU Mic in the quiescent state. Since there are no available flaring spectra data in the XUV and FUV range for AU Mic, we added this scaling factor to act as a baseline for the emission of AU Mic during flares. \citet{2022Feinstein} recently provided the AU Mic spectra composed of observations in the quiescent state, along with stellar spectra models. The observations were done using the XMM-Newton, covering 0.1-3.9 nm, and FUSE, from 90 - 118.1 nm. The DEM model was used to fill out the gap between 4 and 90 nm with a synthetic spectrum. These composed spectra give us fluxes in the FUV and XUV range, which we use to calculate the scaling factor of XUV/FUV, resulting in  $\approx$2.182. The X-ray emission from AU Mic in the quiet state makes up 11$\%$ of the XUV, which is comparable to the 15$\%$ found by the MUSCLES program using a M1V star. 

These two scaling factors allow us to account for the role of systematic uncertainties by accessing two complementary sources: emission from flares and emission from the AU Mic star. While the calculation of the first scaling factor is based directly on flare emission but is not related to AU Mic itself, the second scaling factor provides information about the lower limits of flaring emission that AU Mic should exhibit in the XUV range.

Converting the flare energies from FUV to XUV implies that the FFD fit will change due to the new energy values in the XUV bandpass. Considering that the slope $\alpha$ stays the same in both bandpasses \citep{maehara2015statistical,jackman2023extending}, only the FFD Y-Intercept in the XUV bandpass ($\beta_{XUV}$) has to be scaled. According to \citet[Eq. 5]{jackman2023extending}, the $\beta_{XUV}$ is

\begin{align*} \label{eq:beta_euv}
    \beta_{XUV} = \beta_{FUV} -\alpha log10(f) \,
\end{align*}

where $f$ is the ratio between the flux of the band and the FUV.

\begin{table*}[t]
\centering
\caption{Scaling factors from FUV to XUV bandpass.}
\label{tab:convert}
\begin{tabular}{ccc}
\hline \hline
Source               &  XUV/FUV   & Scaling Factor Name \\ \hline
MUSCLES panchromatic spectra in flaring state $^a$     &  2.8  & With Flares (factor 2.8)    \\
AU Mic panchromatic spectra in quiescent state $^b$ &   2.2 & With Flares (factor 2.2)\\
\hline
\end{tabular}
\begin{flushleft}
\footnotesize{$^a$ GJ 832 flaring spectra from the MUSCLES project \citet{2018Loyd}.}\\
\footnotesize{$^b$ Derivate from the AU Mic emission spectra using XMM-Newton, FUSE observations, and the DEM model \citep{2022Feinstein}.} 
   
\end{flushleft}

\end{table*}

\vspace{1.5cm}

\subsection{Atmospheric Escape}\label{sec:atmesc}

To calculate the hydrogen loss, we use the \atmesc module from \vplanet, which is responsible for the loss processes in the planet's atmosphere. For this work, we use two approaches to calculate the atmospheric escape of the AU Mic planets: the radiation-recombination regime \citep{murray2009atmospheric} and the energy-limited regime \citep{erkaev2007roche,luger2015habitable}. \vplanet switches automatically between the two regimes during the simulations depending on the incoming flux value at the planet. These processes are two special cases of hydrodynamic escape. In summary, at low XUV flux (F$_{XUV}$ $\lesssim$ 10 W/m$^{2}$), the hydrodynamic escape follows the energy-limited regime. In this case, the low flux allows a larger portion of the incoming radiation to heat the gas. This heating results in the gas expanding and cooling down (PdV work), rather than having the energy lost to radiation from ion-electron recombinations, and the planet will lose its atmosphere at a mass rate $\dot{M}_{EL}$ given by 

\begin{equation}
\dot{M}_{EL} = \frac{\epsilon_{XUV}\pi F_\mathrm{XUV}R_\mathrm{p}R_\mathrm{XUV}^{2}}{G M_\mathrm{p}K_\mathrm{tide}},
\label{eq:EL}
\end{equation}
where $\epsilon_{XUV}$ is the hydrogen escape efficiency due to the XUV (varying from 0 to 1;  \citealt{barnes2020vplanet}), $F_\mathrm{XUV}$ is the XUV flux incoming to the planet \citep{erkaev2007roche}, $M_\mathrm{p}$ is planetary mass, and $G$ is the gravitational constant. The $K_\mathrm{tide}$ is the parameter that represents the tidal effect in the atmosphere, and its typical values go from 0.9 to 0.99 for terrestrial planets in the HZ of M stars \citep{luger2015habitable}.

On the other hand, at high XUV flux (F$_{XUV}$ $\gtrsim$ 10 W/m$^{2}$), the increasing fractions of the energy hitting the upper atmosphere will be lost to ion-electron recombinations, and the hydrodynamic escape will be mainly in the radiation-recombination regime, with the mass loss being driven more slowly \citep{murray2009atmospheric}. If the flux is equal to or higher than the flux threshold $F_{crit}$ given by

\begin{equation}
    F_{crit} = \Big(\frac{A}{B}\Big)^2,
    \label{eq:rrcrit}
\end{equation}
where
\begin{equation}
    A = 2.248\times 10^6\Big(\frac{R_p}{R_\oplus}\Big)^{3/2} \textrm{kg}^\frac{1}{2} s^\frac{1}{2},
    \label{eq:rrB}
\end{equation}
and
\begin{equation}
    B = \frac{\epsilon_{XUV} \pi R_{XUV}^3}{GM_pK_{tide}}.
    \label{eq:rrA}
\end{equation}
Thus, the mass loss rate $\dot{M}_{RR}$ through this escape process is then given by
\begin{equation}\label{eq:RR}
        \dot{M}_{RR} = 2.248 \times 10^6 (F_\mathrm{XUV}R_{p}^{3})^{1/2},
\end{equation}
where $F_\mathrm{XUV}$ is the incoming flux on the planet in the XUV wavelength range \citet[Section 2]{murray2009atmospheric}, $R_{p}$ is the planetary radius, and $R_{XUV}$ is the XUV radius of influence, measured from the center of the planet, which indicates how much the XUV can penetrate into the atmosphere of the planet.

\subsection{Simulations} 

Using the modules \stellar, \atmesc, and \flare from \vplanet, we performed three scenarios of simulations. The first two simulation cases considered the XUV flare contribution in the AU Mic system evolution (using the scaling factors from Table \ref{tab:convert}). The third scenario was done without flares to serve as a case control, i.e., only the quiescent emission of the star was taken into consideration. Even though simulating a case without flares is unrealistic, it is widely used to calculate the evolution of the stellar flux reaching the planets, since the quiescent emission is the baseline and makes up the major part of the stellar emission over the lifetime of a star, regardless of the spectral type. This assumption is broadly made because, while is true that M stars can produce stellar flares whose emission can increase the stellar flux by two orders of magnitude over the quiet flux \citep{scalo2007m}, this type of event lasts only for a short time scale of minutes to a few hours. By considering simulations with and without the extra XUV due to stellar flaring, we can quantify how much of the atmosphere of the planets in the AU Mic system is lost due to the quiescent emission and how much is from the influence of stellar flares.

Since all the AU Mic planets have published constraints on their masses and radii, we use this information as boundary conditions. By using these values, we ensure that during the simulated system's evolution, the planets will match the observed mass and radii at the current age of the system, which is about 23 Myrs. In our simulations, we worked backward to find the necessary input values (at 10 Myrs) for the planets' masses and their hydrogen envelopes that will match the present planetary mass and radius values. This methodology allows us to account for the loss of atmosphere that has occurred as the planets are evolving to their current state, thus ensuring that the final values for mass and radius match the present values. Another important point is that all the simulations start at 10 Myrs, which is the stellar age at which the protoplanetary disk has mostly dissipated and the planets are formed \citep{Raymond07,Lambrechts19,Clement22}. Table \ref{tab:simulations} presents the values for the parameters used as boundary conditions (at 23 Myrs) and input parameters (at 10 Myrs). 

Several previous papers have found the value for the masses and radii of AU Mic b and c through independent observation campaigns, using radial velocity and transit time variations (TTVs) measurements \citep{plavchan2020planet,2021Cale,2021Klein,2021Martioli,2022Gilbert,szabo2022transit,2023Donati,2023WittrockAUMic,zicher2022one}. Aiming to increase the accuracy of mass and radius for the planets in the AU Mic system, \citet{mallorquin2024revisiting} combined multiple data of past observations, using almost 500 high-resolution spectra for radial velocity measurements, along with transit detections from the Transiting Exoplanet Survey Satellite (TESS; \citealt{ricker2015transiting}) and CHaracterising ExOPlanet Satellite (CHEOPs; \citealt{rando2018cheops}). \citet{mallorquin2024revisiting} found that AU Mic b should have a mass and radius of 8.99$^{+2.61}_{-2.67}$ $M_{\oplus}$ and 4.79$^{+0.29}_{-0.29}$ $R_{\oplus}$, and AU Mic c of 14.46$^{+3.24}_{-3.42}$ $M_{\oplus}$ and 2.79$^{+0.18}_{-0.17}$ $R_{\oplus}$. Considering the large amount of RV collected by \citet{mallorquin2024revisiting}, we chose their results for the planetary mass and radius of AU Mic b and c as boundary conditions at 23 Myrs in the simulations.

On the other hand, \citet{2023WittrockAUMic} confirmed AU Mic d using TTVs, showing that AU Mic d is an Earth-sized planet and has an estimated mass of 1.053 M$_{\oplus}$ and a radius of 1.023 R$_{\oplus}$. Since the results from \citet{2023WittrockAUMic} are the only predicted values for AU Mic d until this date, we use them as boundary conditions for the simulations of AU Mic d. We address the uncertainties of mass and radius from the planets by performing simulations considering the upper and lower limit errors (see the last two rows of Table \ref{tab:simulations} for the uncertainty values). The results for accounting errors are available in Table \ref{tab:results2} and are represented as a low opacity shadow in Figure \ref{fig:atm}. All the results and figures from this work are available on a GitHub repository\footnote{\url{https://github.com/lauraamaral/AUMicAtmEvolution.git}}.

\begin{scriptsize}
\begin{table*}[t]
\caption{Input parameter space and boundary conditions considered in the simulations.}
\label{tab:simulations}
\begin{tabular}{lcccc}
\hline
Stellar Parameter                         & AU Mic            & & & \\  \hline
Stellar Mass ($M_{\odot}$)                & 0.510$^{a}$       & & & \\
XUV Saturation Time (Gyr)                 & 0.149$^{b}$       & & & \\
XUV Saturation Fraction                   & 10$^{-3.07 c}$     & & & \\
Rotation Period (days)                    & 4.856$^{d}$       & & & \\
Flare Energy in FUV (ergs)                & 10$^{28}$ - 10$^{34 e}$ & & & \\ \hline
Initial conditions at 10 Myrs   & & & & \\ \hline
Planetary Parameter                       & XUV Evolution     &AU Mic b    &AU Mic c& AU Mic d \\ \hline
Planetary Mass ($M_{\oplus}$)$^{f}$       & With Flares (factor 2.8) &9($_{-3}^{+3}$)    & 14.46($_{-3.42}^{+3.24}$) & 1.054($_{-0.511}^{+0.511}$)  \\
                                          & With Flares (factor 2.2) &9($_{-3}^{+3}$)    & 14.46($_{-3.42}^{+3.24}$)  & 1.054($_{-0.511}^{+0.511}$)   \\
                                          & No Flares &8.99($_{-2.66}^{+2.61}$) & 14.46($_{-3.42}^{+3.24}$)  & 1.054($_{-0.511}^{+0.511}$)  \\\hline
Planetary Radius ($R_{\oplus}$)$^{f}$       &With Flares (factor 2.8)  &5.35($_{-0.22}^{+0.23}$) & 2.89($_{-0.17}^{+0.18}$)  & 1.42($_{-0.09}^{+0.36}$)  \\
                                          & With Flares (factor 2.2) &5.35($_{-0.22}^{+0.23}$) & 2.89($_{-0.17}^{+0.18}$)  & 1.41($_{-0.08}^{+0.34}$)   \\
                                          & No Flares                &5.35($_{-0.22}^{+0.23}$) & 2.89($_{-0.17}^{+0.18}$)  & 1.39($_{-0.06}^{+0.29}$)  \\\hline
Envelope Mass ($M_{\oplus}$)$^{f}$        & With Flares (factor 2.8) & 0.79($_{-0.35}^{+0.49}$)   & 0.21($_{-0.089}^{+0.12}$)  & 1.08($_{-0.18}^{+0.3}$)$\times$ 10$^{-3}$  \\
                                          & With Flares (factor 2.2) & 0.79($_{-0.35}^{+0.49}$)   & 0.21($_{-0.09}^{+0.12}$)   & 1.04($_{-0.18}^{+0.29}$)$\times$ 10$^{-3}$  \\
                                          & No Flares & 0.79($_{-0.34}^{+0.49}$)   & 0.21($_{-0.09}^{+0.12}$)   & 9.1($_{-0.21}^{+0.252}$)$\times$ 10$^{-4}$  \\\hline
Eccentricity$^{a}$                        & All                & 0.07      & 0.18     &  0.003  \\
Orbital Period$^{a}$ (days)               & All                & 8.464  &18.86 & 12.74  \\
Thermosphere Temperature$^{g}$ (K)        & All & 554.8 & 424.7 & 143.4$^{h}$ \\    
XUV Hydrogen Escape Efficiency$^{i}$      & All & 0.15  & 0.15  & 0.15 \\
Simulation Time (Gyr)$^{j}$               & All & 5     & 5     &  5   \\
Time step (yr)                            & All & 1-10$^{4}$ & 1-10$^{4}$ & 1-10$^{4}$ \\ \hline
Boundary conditions at 23 Myrs  & & & & \\ \hline
Planetary Radius ($R_{\oplus}$)$^{k}$      & All & 4.79$_{-0.29}^{+0.29\,l}$ & 2.79$_{-0.17}^{+0.18\,l}$ & 1.023$_{-0.139}^{+0.139\,a}$ \\
Planetary Mass ($M_{\oplus}$)             & All & 8.99$_{-2.67}^{+2.61\,l}$ &14.46$_{-3.42}^{+3.24\,l}$ &1.053$_{-0.511}^{+0.511\,a}$ \\\hline
\end{tabular}
\begin{flushleft}
\footnotesize{$^a$ From \citet{2023WittrockAUMic}.}\\
\footnotesize{$^b$ From \citet{richey2023hazmat}.}\\
\footnotesize{$^c$ From \citet{wright2011stellar}.}\\
\footnotesize{$^d$ From \citet{2023Donati}.}\\
\footnotesize{$^e$ The FUV observations from HAZMAT program show energies up to 10$^{32}$ ergs. However, AU Mic has an observed FFD in the NUV (303–417 nm, \citet{tristan20237}) with energies up to 1.6 $\times$ 10$^{33}$ ergs. Using the \citet{jackman2023extending} scaling factor from NUV to FUV, we found that these NUV flares have 2.4 $\times$ 10$^{33}$ ergs in the FUV range, which we approximate to 10$^{34}$ ergs for the purpose of this work}.\\
\footnotesize{$^f$ This work.}\\
\footnotesize{$^g$ From \citet{mallorquin2024revisiting}.}\\
\footnotesize{$^h$ Since AU Mic d has no constraints as to the temperature, we calculated the temperature using the Stefan-Boltzmann law. The value in the table represents the value at the beginning of the simulation, at 10 Myr. However, this value changes over time.}\\
\footnotesize{$^i$ From \citet{barnes2020vplanet}.}\\
\footnotesize{$^j$ The field age from the stars of the MUSCLES project goes up to 9 Gyrs. However, its lower limit goes up to 5 Gyrs. For reliable results, we simulated the system for 5 Gyrs instead.}\\
\footnotesize{$^k$ To simulate the evolution of the planetary radii of AU Mic d, we used the planetary radius evolutionary model from \citet{luger2015habitable}, which is a modification from \citet{lopez2012thermal} to better fit terrestrial planets. For all the other planets, we use the planetary radius evolutionary model from \citet{lopez2012thermal}.}\\
\footnotesize{$^l$ From \citet{mallorquin2024revisiting}.}\\   
\end{flushleft}
\end{table*}
\end{scriptsize}

\newpage
\section{Results} \label{sec:results}

\begin{center}
\begin{scriptsize}
\begin{table*}[t]
\centering
\caption{Simulation results for the planets AU Mic b, c, and d, using the synthetic FFDs built with the scaling factors from Table \ref{tab:convert}.}
\label{tab:results2}
\begin{tabular}{lcccc}
\hline
Parameter                                  &  XUV Evolution   & AU Mic b                 & AU Mic c                 & AU Mic d   \\ \hline
\hline 
Results at 23 Myrs   &&&& \\ \hline 

Escaping rate (g/s)                        & With Flares (factor 2.8) &1.28($_{-0.12}^{+0}$) $\times$ 10$^{11}$ & 5.01($_{-0.43}^{+0.07}$) $\times$ 10$^{9}$ & 5.53($_{-0}^{+0.07}$) $\times$10$^{9}$\\ 
                                           & With Flares (factor 2.2) &1.26($_{-0.11}^{+0}$) $\times$ 10$^{11}$ & 4.81($_{-0.42}^{+0.07}$) $\times$ 10$^{9}$ & 5.30($_{-0.71}^{+0.03}$) $\times$ 10$^{9}$  \\
                                           &             No Flares &1.08($_{-0.04}^{+0.09}$) $\times$ 10$^{11}$ & 4.01($_{-0.35}^{+0.06}$) $\times$ 10$^{9}$ & 4.42($_{-0}^{+0.06}$) $\times$10$^{9}$\\ \hline                                           
Hydrogen Envelope Mass (M$_{\oplus}$)      & With Flares (factor 2.8) & 0.78($_{-0.34}^{+0.49}$)  & 0.21($_{-0.09}^{+0.12}$)  & 1.05($_{-0.53}^{+3.99}$) $\times$ 10$^{-5}$ \\
                                           & With Flares (factor 2.2) & 0.78($_{-0.34}^{+0.49}$)  & 0.21($_{-0.09}^{+0.12}$)  & 1.3($_{-0.64}^{+5.04}$) $\times$ 10$^{-5}$ \\
                                           &                No Flares & 0.78($_{-0.34}^{+0.49}$)  & 0.21($_{-0.09}^{+0.12}$)  & 1.04($_{-0.57}^{+3.99}$) $\times$ 10$^{-5}$ \\ 
\hline
Results at 5 Gyrs     &&&& \\ \hline 

Escaping rate (g/s)                        & With Flares (factor 2.8) & 2.74($_{-0.06}^{+0.24}$) $\times$ 10$^{8}$ & 2.53($_{-0.003}^{+0.17}$) $\times$ 10$^{7}$ & 0 \\ 
                                           & With Flares (factor 2.2) & 2.62($_{-0.05}^{+0.22}$) $\times$ 10$^{8}$ & 2.41($_{-0.003}^{+0.16}$) $\times$ 10$^{7}$ & 0 \\
                                           & No Flares                & 2.13($_{-0.006}^{+0.16}$) $\times$ 10$^{8}$ & 1.92($_{-0.003}^{+0.13}$) $\times$ 10$^{7}$ & 0 \\  \hline
                                        
Hydrogen Envelope Mass (M$_{\oplus}$)      & With Flares (factor 2.8) &0.68($_{-0.35}^{+0.49}$)  & 0.2($_{-0.09}^{+0.12}$)  & 0\\
                                           & With Flares (factor 2.2) &0.69($_{-0.35}^{+0.49}$)  & 0.2($_{-0.09}^{+0.12}$)  & 0 \\
                                           & No Flares  &0.71($_{-0.35}^{+0.49}$)  &  0.2($_{-0.09}^{+0.12}$)  & 0\\ \hline
                                           
Planetary Radius (R$_{\oplus}$)            & With Flares (factor 2.8) &3.17($_{-0.38}^{+0.38}$)  & 2.46($_{-0.16}^{+0.17}$)  & 1.014($_{-0.168}^{+0.115}$) \\
                                           & With Flares (factor 2.2) &3.18($_{-0.37}^{+0.37}$)  & 2.46($_{-0.16}^{+0.17}$)  & 1.014($_{-0.168}^{+0.115}$) \\
                                           & No Flares &3.2($_{-0.36}^{+0.37}$)  & 2.46($_{-0.16}^{+0.17}$)  & 1.014($_{-0.168}^{+0.115}$) \\ \hline   
                                           
Planetary Mass (M$_{\oplus}$)              & With Flares (factor 2.8) &8.89($_{-2.67}^{+2.61}$)  & 14.45($_{-3.42}^{+3.24}$)  & 1.053($_{-0.511}^{+0.511}$) \\
                                           & With Flares (factor 2.2) &8.9($_{-2.67}^{+2.61}$)  & 14.45($_{-3.42}^{+3.24}$)  & 1.053($_{-0.511}^{+0.511}$) \\
                                           & No Flares &8.92($_{-2.66}^{+2.61}$)  & 14.46($_{-3.42}^{+3.24}$)  & 1.053($_{-0.511}^{+0.511}$) \\ \hline 
                                        
Planetary Mean Density (g/cm$^{3}$)        & With Flares (factor 2.8) &1.54 & 5.35 & 5.57\\
                                           & With Flares (factor 2.2) &1.53 & 5.35 & 5.57\\ 
                                           & No Flares &1.5 & 5.34 & 5.57\\ \hline                                           
\end{tabular}
\end{table*}
\end{scriptsize}
\end{center}

\subsection{Escape Rates of AU Mic planets}

Even though all three planets are located between the host star and the inner limit of the conservative habitable zone (HZ), only AU Mic d is calculated to have a total depletion of its primordial atmosphere, which happens by about $\sim$ 20 Myrs (see Figure \ref{fig:atm}.b). Considering the uncertainty in the AU Mic's age, AU Mic d may have already lost its primordial atmosphere, or it could be about to happen.

To fit the planetary mass and radius values of AU Mic d found by \citet{2023WittrockAUMic}, the current hydrogen envelope mass (i.e., at 23 Myrs) should be around 1.04-1.05 $\times$ 10$^{-5}$ M$_{\oplus}$ (see Table \ref{tab:results2}), which is equivalent to $\approx$12 times the Earth's atmospheric mass. That means AU Mic d should have started its evolution with a primordial atmosphere of almost 10$^{-3}$ M$_{\oplus}$. The current estimated escape rate found in the simulations for AU Mic d is $\approx$ 4.4 - 5.5 $\times$ 10$^{9}$ g/s.

\begin{figure}
    \centering
    \includegraphics[width=0.999\linewidth]{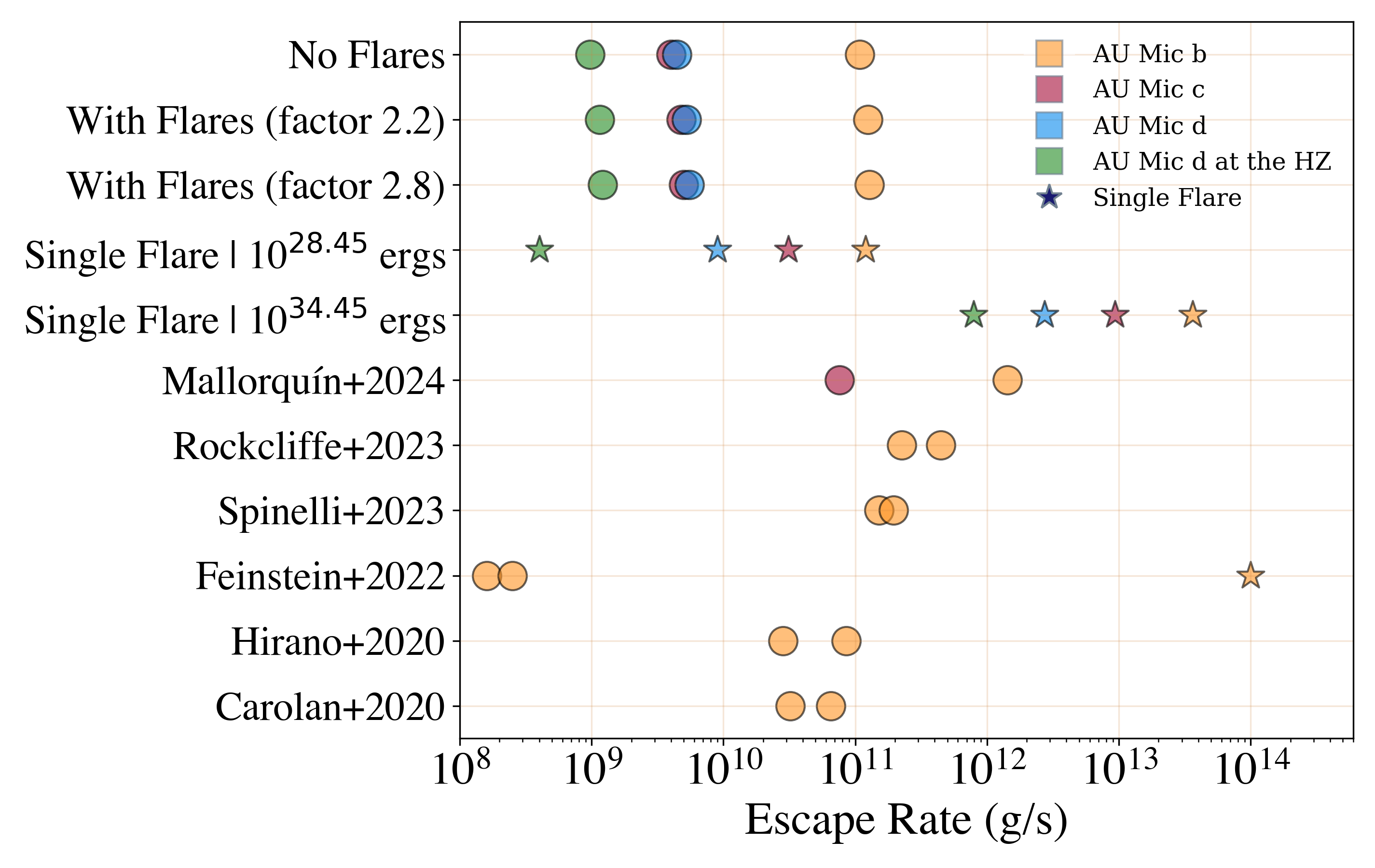}
    \caption{Escape rates from AU Mic b, c, and d calculated in this work and compared to past work. The star symbols represent the instantaneous escape rate driven for a single flare. The single flare energies are calculated using the scaling factor XUV/FUV = 2.8.}
    \label{fig:escape_rates}
\end{figure}

\begin{table*}[t]
 \caption{Instantaneous escape rates (g/s) driven by a single flare for the AU Mic planets.}
    \label{tab:singleflareescaperates}
    \centering
    \begin{tabular}{cccccc} \hline
        XUV Flare Energy$^{a}$ (erg) & FUV Flare Energy (erg)  & AU Mic b & AU Mic c & AU Mic d & AU Mic d at the HZ$^b$ \\ \hline
        10$^{28.45}$  & 10$^{28}$  & 1.20$\times$10$^{11}$ & 4.08$\times$10$^{9}$ & 9.02$\times$10$^{9}$ & 4$\times$10$^{8}$\\
        10$^{34.45}$  &10$^{34}$  & 3.62$\times$10$^{13}$ & 9.36$\times$10$^{12}$ & 2.72$\times$10$^{12}$ & 7.91$\times$10$^{11}$ \\ \hline
    \end{tabular}
    \begin{flushleft}
        \footnotesize{$^a$ The single flare energy in the XUV is calculated using the scaling factor XUV/FUV = 2.8.}\\
        \footnotesize{$^b$ AU Mic d located at the Earth equivalent position from the Sun, 0.2935 AU, receiving an XUV flux of 1.44 W/$m^{2}$.}\\
    \end{flushleft}
\end{table*}

In contrast to AU Mic d, the planets AU Mic b and c do not have their atmospheres fully depleted. The results (see Table \ref{tab:results2}) show that for AU Mic b to have its currently measured mass and radius, it should have an atmosphere of $\approx$ 0.779 M$_{\oplus}$. Even though AU Mic b has the largest escape rate (1.1-1.3 $\times$ 10$^{11}$ g/s) of all the planets in the system, it does not lose its entire primordial atmosphere, not even after 5 Gyrs of evolution. This might happen because, despite the proximity of the host star, AU Mic b is a sub-Neptune planet with a higher atmospheric reservoir than AU Mic d. AU Mic c presents a similar behavior. We found that AU Mic c might currently have an atmospheric mass of $\approx$ 0.208 M$_{\oplus}$, and the lowest escape rate of the three planets: $\approx$ 4-5 $\times$ 10$^{9}$ g/s. This is because AU Mic c is a sub-Neptune with a gravity higher than AU Mic d, and it is located farther away from the host star, where the lower XUV flux reaching the planet (about $\sim$ 10 W/m$^{2}$ at 23 Myrs, 3$\times$ lower than the fluxes received by AU Mic d) is not enough to fully deplete its atmosphere.

All these values are calculated based on the quiescent emission plus the average XUV flux emitted by flares per time step. To understand the effect of a single flare on the atmospheric loss rate of the AU Mic planets, we also calculate the instantaneous escape rate driven by a single flare using the respective lower and upper limit flare energies used in the simulations (star symbols in Figure \ref{fig:escape_rates}; Table \ref{tab:singleflareescaperates}). These results show that the escape rates for AU Mic b can go from 1.20 $\times$ 10$^{11}$ g/s under a single flare of 10$^{28}$ ergs to 3.62 $\times$ 10$^{13}$ g/s for a superflare of 10$^{34}$ ergs, which is about one order of magnitude lower than the value of 10$^{14}$ g/s found by \citet{2022Feinstein} for a flare with energy $\sim$ 10$^{36}$ ergs.

For AU Mic b and d, the escape rate increases about 100 times over the quiescent emission levels when the planets are hit by a single FUV flare of 10$^{34}$ ergs. For AU Mic c, this value goes over 10$^{4}$ the quiescent level. On the other hand, the flare with the lowest energy considered in our simulations increases the escape rate no more than 8 times the quiescent levels for all the planets.

When we repositioned AU Mic d at 0.2935 AU (i.e., 1 AU equivalent at the HZ), the escape rates dropped significantly, presenting values between 9.7 $\times$ 10$^{8}$ g/s without flares and 1.2$\times$ 10$^{9}$ g/s with flares. These escape rates imply that flares will accelerate the escape process by over a thousand years. In section \ref{sec:flarinfaffect}, we use AU Mic d as a case study to understand how the XUV flux of flares affect terrestrial planets at different distances from the host star.

\begin{figure*}[t!]
\includegraphics[width=\textwidth]{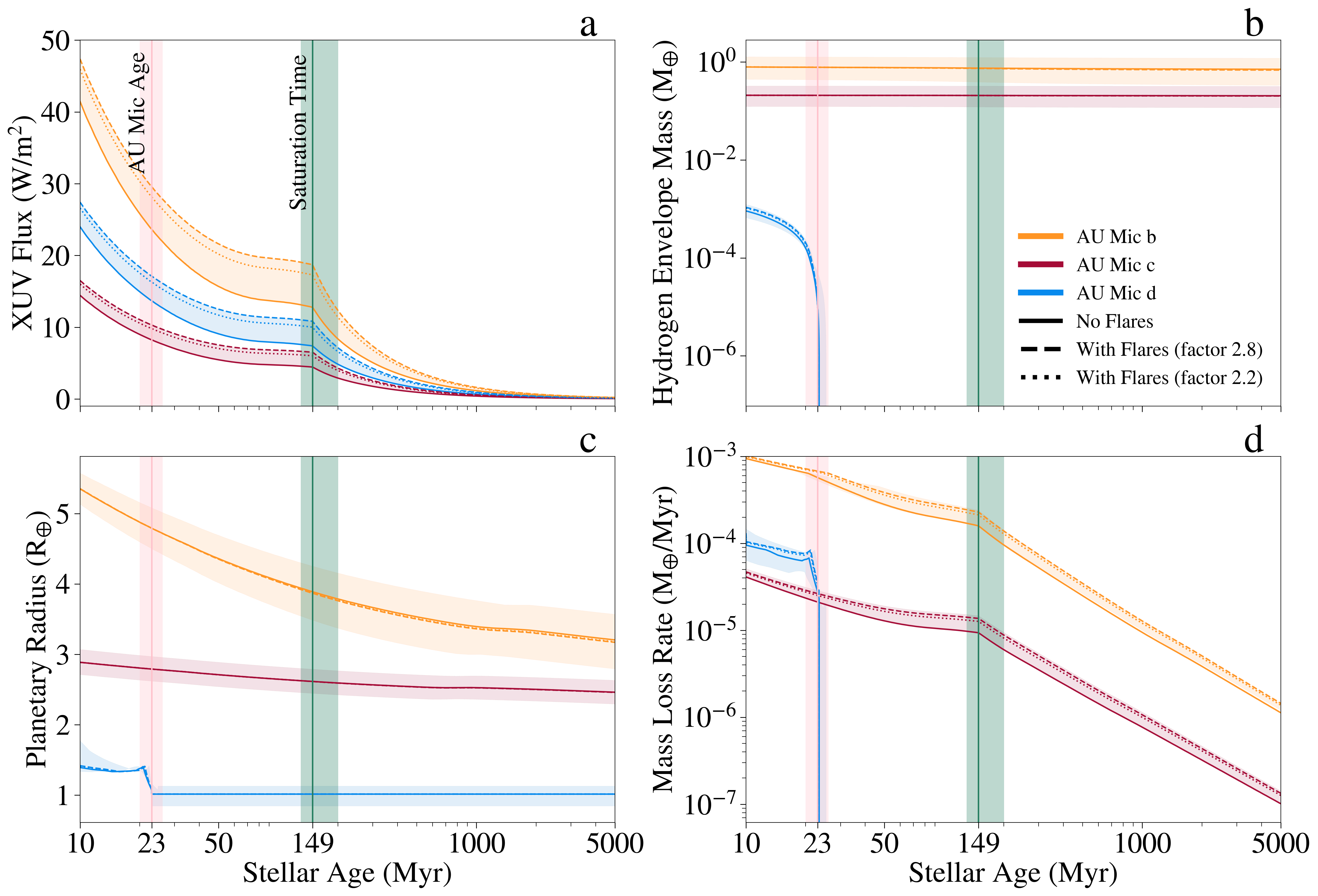}
\caption{Atmospheric evolution of the planets AU Mic b (closest from the star), d (in between AU Mic b and c), and c (furthest from the star). a) XUV flux reaching the planet, b) primordial H/He envelope mass, c) planetary radius, and d) loss rate of the atmospheric mass. The shadows represent the associated errors in the results, considering the uncertainties for the current masses and radii of the AU Mic planets.}
\label{fig:atm}
\end{figure*}
 
\subsection{Past, present and future of AU Mic planets}

\begin{figure}[htp!]
\centering
 \includegraphics[width=1\linewidth]{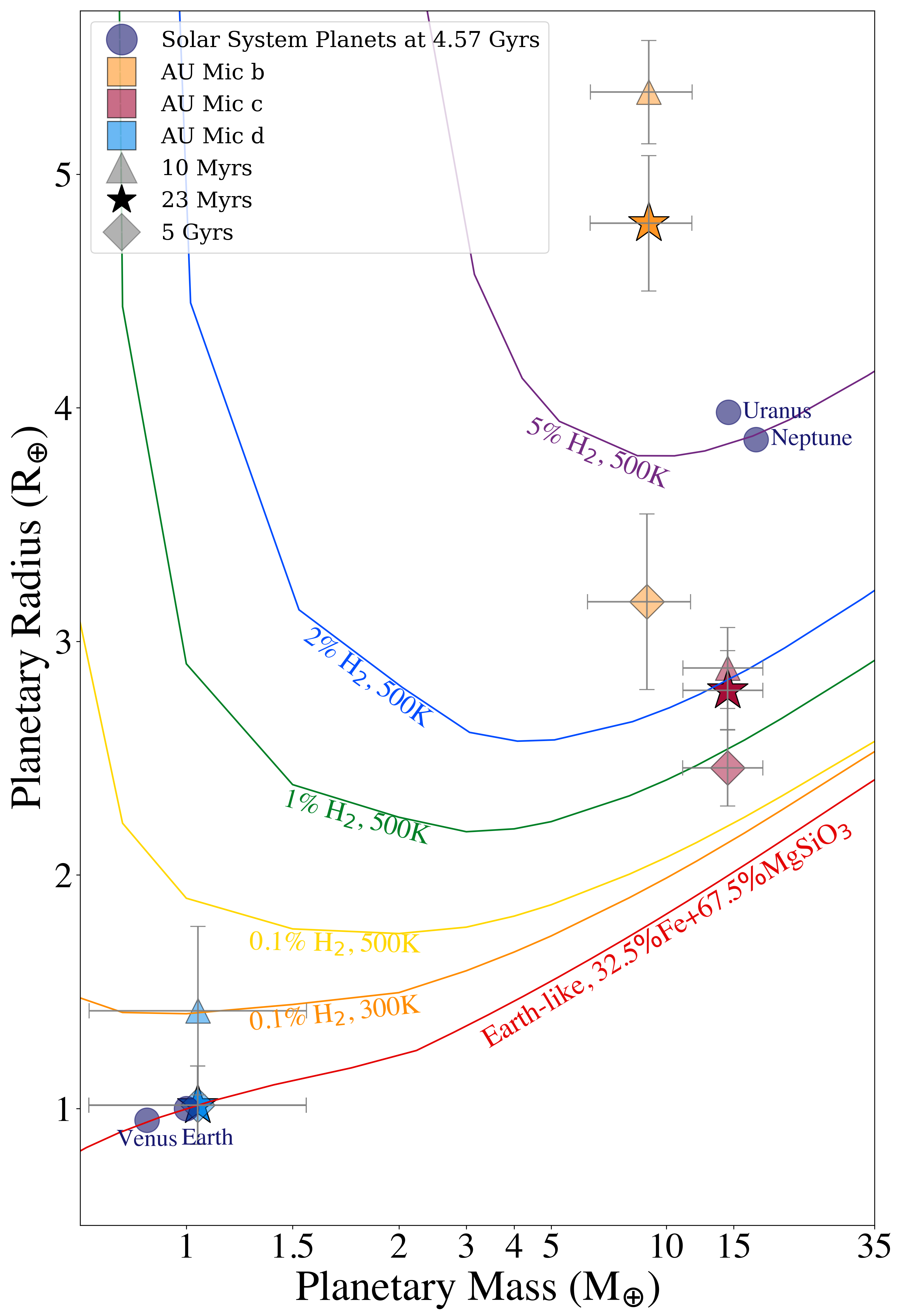}
\caption{Mass-radius diagram for the AU Mic planets. The solid lines represent the mass-radius relationships for Earth-like cores (32.5$\%$ Fe core and 67.5$\%$ MgSiO$_{3}$ mantle) with different amounts of hydrogen in the atmosphere \citep{2016Zeng,zeng2019growth}. The results for AU Mic b (yellow markers), c (red markers), and d (blue markers) at the system age of 10 Myrs, 23 Myrs, and 5 Gyrs are represented by triangles, stars, and diamonds, respectively. Venus, Earth, Uranus, and Neptune are also shown in the figure, represented by dark purple circles.} \label{fig:m-r}
\end{figure}

All the planets' radii shrink over time due to escape and atmospheric cooling since the beginning of the evolution at 10 Myrs (see Figure \ref{fig:atm}.b and c). To match the observed data at the age of 23 Myrs and compensate for the atmospheric loss over time, the planets need to start their evolution (at 10 Myr) with a certain ratio of atmosphere to solid core. Also, the exact amount of atmospheric hydrogen mass and solid core mass will change depending on how much atmosphere was lost over time due to the XUV flux. This implies that we can estimate not only the future of the AU Mic planets but also the state of these planets at the beginning of their evolution. These results can help us understand how the process of atmospheric loss by planets changes their size over time and the role of stellar flares in this evolution. Figure \ref{fig:m-r} shows a mass-radius relationship diagram for the AU Mic planets and how their potential composition changes due to atmospheric loss over different ages. The masses and radii of the three AU Mic planets are shown at 10 Myrs (triangle markers), 23 Myrs (star markers), and 5 Gyrs (diamond markers), as well as which composition model would fit these values (solid lines). The models used here to fit the potential composition of the AU Mic planets are based on the mass-radius relationships from \citet{2016Zeng} and \citet{zeng2019growth}.

\subsection{How do stellar flares affect atmospheric loss of planets?}\label{sec:flarinfaffect}

The results for the AU Mic planets do not show profound differences between the impact of the quiescent emission alone and the cases where we add flares. The escape rates at 23 Myrs increase by approximately 15$\%$ due to flaring but are not enough to promote a faster atmospheric depletion when compared to quiescent emission. Figure \ref{fig:atm} shows that XUV flux increases by 6 W/m$^{2}$ for the closest planet (AU Mic b) when stellar flares are added, which represents 15$\%$ of the total XUV flux.

The simulation results show that the XUV contribution from stellar flaring increases from about 10$\%$ at the beginning of the evolution to roughly 30$\%$ of the total XUV emission from the star after the end of the saturation phase (after $\sim$ 150 Myrs, see vertical green line in Figure \ref{fig:star}). This result means that the quiescent emission levels are still enough to dominate the XUV emission in the early and late stages of the star evolution, even when the flaring contribution reaches its maximum around the end of the saturation phase.

Another feature that influences this result is the distance from the host star. All three AU Mic planets are located between the host star and the HZ inner edge, placing them out of the HZ and close-in to the host star, which can expose them to high levels of XUV radiation. To understand how the distance and flux affect planetary atmospheric retention, we ran simulations placing AU Mic d at different distances from the host star (see Table $\ref{tab:distance}$).

\begin{figure}[!]
    \centering
    \includegraphics[width=1\linewidth]{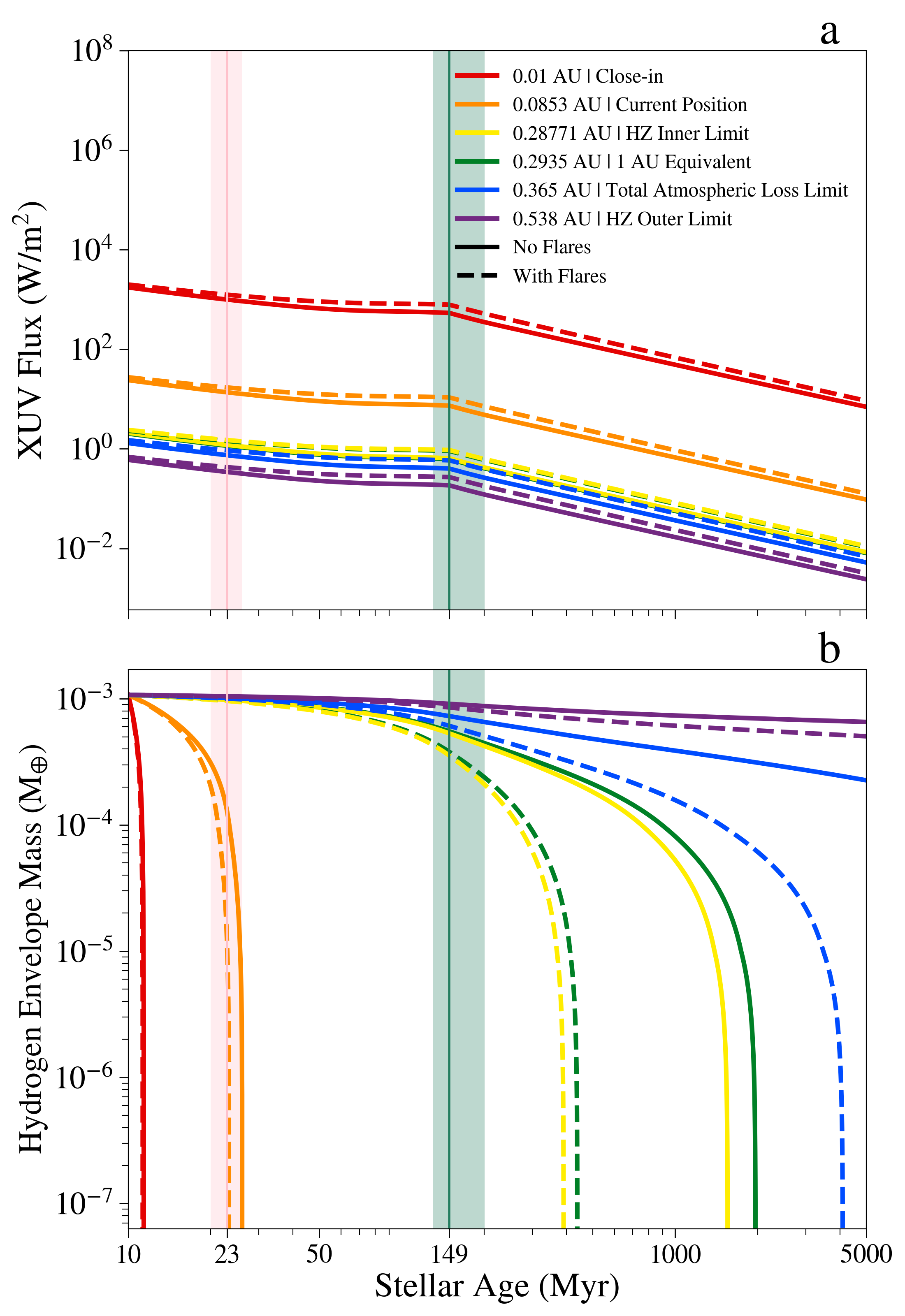}
    \caption{The atmospheric evolution of AU Mic d at different distances to the host star. The simulations considering flares in the evolution are represented by dashed lines, while simulations with quiescent levels of XUV are represented by solid lines. The different colors represent AU Mic d placed close-in to the star (i.e., at 0.01 AU; red lines), at the current position (i.e., at 0.0853 AU; orange lines), at the inner limit of the HZ (i.e., at 0.28771 AU; yellow lines), and at the outer limit of the HZ (i.e., at 0.538 AU; violet lines). The green lines represent AU Mic d at the same equivalent distance of the Earth inside the HZ (i.e., 0.2935 AU). The blue lines represent the furthest position that AU Mic d can be for the total atmospheric loss to happen (i.e., 0.365 AU).}
    \label{fig:AUMicdDistance}
\end{figure}

The results from Figure \ref{fig:AUMicdDistance} show AU Mic d placed close-in to the host star (i.e., at 0.01 AU; red lines), at the current distance (i.e., at 0.0853 AU; orange lines), and at the inner (i.e., at 0.28771 AU; yellow lines) and outer edge of the HZ (i.e., at 0.538 AU; violet lines). We also placed AU Mic d at the furthest position where the total atmospheric loss of AU Mic can still happen, before not even the extra XUV flux from stellar flares is enough to fully deplete AU Mic d's atmosphere (i.e., 0.365 AU; blue lines). These distances resulted in a flux of approximately 1244, 17.1, 1.5, 0.43, and 0.934 W/m$^{2}$, respectively, as shown in Figure \ref{fig:AUMicdDistance}.a. We also plot AU Mic d inside the conservative HZ at the same equivalent Earth distance from the Sun (i.e., 1 AU equivalent, green lines), which results in an XUV flux of 1.44 W/m$^{2}$ reaching the planet. This value is in agreement with the results found by \citet{richey2023hazmat}, which estimate that planets at the HZ of M stars will receive X-ray fluxes between 0.848 and 1.85 W/m$^{2}$ and FUV fluxes between 0.064 and 0.038 W/m$^{2}$.

\begin{table}
 \caption{Input AU Mic d sintetic positions considered in Figure $\ref{fig:AUMicdDistance}$.}
    \label{tab:distance}
    \centering
    \begin{tabular}{ccc} \hline
        AU Mic d Position   & Distance (AU) & XUV Flux$^{a}$ (W/m$^{2}$)\\ \hline\hline
        Close-in &  0.01 & 1244\\\hline
        Current position & 0.0853 & 17.1 \\\hline
        HZ inner limit & 0.28771 &  1.5 \\\hline
        1 AU equivalent$^{b}$ & 0.2935 & 1.44\\\hline
        Total Atmospheric\\ Loss Limit$^{c}$ & 0.365 &  0.934 \\\hline
        HZ outer limit & 0.538 &  0.43 \\ \hline
    \end{tabular}
    \begin{flushleft}
     \footnotesize{$^a$ XUV Flux at 23 Myrs considering the quiescent emission plus flares.}\\
     \footnotesize{$^b$ The same equivalent distance of the Earth inside the conservative HZ around the Sun.}\\
     \footnotesize{$^c$ Furthest position that AU Mic d can be and still experience a total depletion of its atmosphere. In this case, the position located at 0.365 AU from the host star represents where the complete depletion of the atmosphere is possible only with the addition of stellar flares. After that point, the flux will not be enough to deplete AU Mic d's atmosphere, even with the extra XUV from flares.}
    \end{flushleft}
\end{table}

The results presented in Figure \ref{fig:AUMicdDistance}.b show that the flare emission will be relevant to the atmospheric escape during the post-saturation phase of early-type Ms, such as AU Mic. For AU Mic d, a terrestrial planet about the size of the Earth, the 15$\%$ increase in XUV flux will not make a difference if the planet is close-in to the star or farther than 0.365 AU. When the planet is too close to the star, the XUV levels are too high ($\approx$ 1000$\times$ the HZ fluxes), in which case the quiescent flux is enough to drive escape and fully deplete the primordial atmosphere of the planet. Similarly, when the planet is too far away, the flux levels are too low ($\leq$ 0.5$\times$ the HZ fluxes), preventing the complete loss of the atmosphere. 
The middle point seems to be when the planet is located inside the HZ, where we can note a clear difference between the impact of the XUV fluxes in quiet and flaring state in the atmospheric evolution (solid lines in Figure \ref{fig:AUMicdDistance}). In this case, when compared to quiescent levels only, flares will accelerate the escape process by a few thousand years. However, XUV flares have the greatest effect on the atmospheric depletion of Earth-sized planets located at 0.365 AU, which for AU Mic is halfway the HZ (blue lines in Figure \ref{fig:AUMicdDistance}). Importantly, in this case, atmospheric depletion will occur only due to the extra XUV from flares. Passing that distance, AU Mic d would no longer have its atmosphere depleted, meaning that Earth-sized planets would be able to keep their atmospheres if they were located on the outer edge of the HZ.

\section{Discussion and Conclusions} \label{sec:conclu}

The close distance to the host star exposes the planets AU Mic b, c, and d to higher levels of XUV (compared to the modern Earth around the Sun), creating an environment that promotes atmospheric escape driven by stellar high-energy radiation. The XUV radiation delivered by stellar flares will also inflate the planetary atmospheres, making the planets more vulnerable to nonthermal escape process \citep{france2020high}. In this work, we simulate the atmospheric evolution of the planets around the young M1V star, AU Mic, exposed to flares by using an analytical approximation that assumes energy-limited escape at low fluxes and transitions to radiation-recombination-limited escape at high fluxes.
We considered the XUV luminosity model from \citet{2005Ribas} for the stellar quiescent XUV evolution, and the fitted grid values of \citet{baraffe2015new} for the evolution of the stellar bolometric luminosity. We also use scaling factors based on spectral data of HAZMAT \citep{loyd2021hazmat} and MUSCLES \citep{2018Loyd} to convert the FFD in the FUV range to XUV and build an FFD age-dependent model. Using this information, we calculate the effect of XUV emission due to stellar flares in the primordial atmospheres of the AU Mic planets over 5 Gyrs of evolution. To calibrate the simulations, we use the measured values of mass and radius for the AU Mic planets \citep{2023WittrockAUMic,mallorquin2024revisiting} as boundary conditions to the simulations.

Our results show that at the age of 23 Myr, the average fluxes received by the AU Mic planets considering the evolution of the XUV emission of stellar flares are up to 29.6, 10.3, and 17.1 W/m$^{2}$, for AU Mic b, c, and d, respectively. These values are 15$\%$ higher than the average XUV fluxes received due to the quiescent emission evolution alone, and $\approx$ 10$^{3}$ times the XUV flux received presently by the Earth \citep[$\approx$ 0.00464 W/m$^{2}$]{luger2015habitable}.

We found that AU Mic d (an Earth-sized planet) is currently likely to be at the limit of completely losing its primordial atmosphere, which our results suggest will happen in a couple of million years from now. Thus, it may be a good target for observing and studying the process of depletion of the primordial atmosphere of a terrestrial planet. We found that AU Mic d should have an escape rate of $\approx$ 5.53 $\times$ 10$^{9}$ g/s at the present day. In agreement to \citet{rockcliffe2023variable}, we also found that AU Mic b and c are in an energy-limited escape regime throughout all 5 Gyrs of evolution, even with the addition of XUV emission of flares. AU Mic b presents the highest escape rate of the AU Mic planets: $\approx$ 1.28 $\times$ 10$^{11}$ g/s. This result is almost 1.5 $\times$ 10$^{11}$ g/s found by \citet{spinelli2023planetary}, and similar to the value calculated by \citet{2020Hirano} of 2.84-8.52 $\times$ 10$^{10}$ g/s. 

The estimates from \citet{2022Feinstein} show that AU Mic b could momentarily reach an escape rate of 10$^{14}$ g/s during the peak of an optical superflare with an energy of $\sim$ 10$^{35.5}$ ergs, and they suggest that flares would dominate the escape of AU Mic planets. On the other hand, our results indicate the opposite, showing that the quiescent emission, not flares, dominates the thermal escape of AU Mic planets in the present day. However, our results were found under the assumption that the flare flux is averaged out over each time step (that varies between days and years), whereas in reality, it will arrive impulsively (in brief spikes). The flux peak values of a flare can temporarily shift the escape regime to a radiation-recombination-limited escape, which is less effective at facilitating atmospheric escape. By averaging the flux, we keep the escape process within the energy-limited regime, which is more efficient. That means the results for the planetary masses, radii, escape rates, and hydrogen envelope mass from this paper could be treated as upper limits. 

For this work, we assumed that the evolution of the flare frequency distribution follows the same saturation-decay relationship that snapshot observations of stellar activity suggest \citep{richey2023hazmat,2024Feinstein}. However, future work is still needed to determine if this is the case for flares in different wavelengths since optical flares seem to have an FFD slope that evolves with time, following a saturation-decay \citep{2024Feinstein}, while FUV FFD slopes seem to vary little in time \citep{loyd2021hazmat}.

We found that during the saturation phase (at high fluxes or close distances), the XUV emission in the quiet state dominates the XUV emission from the star and can drive the depletion of the AU Mic d atmosphere without flares. This indicates that the radius gap found in the Kepler observations \citep{2016Lundkvist,fulton2017california,petigura2022california} is probably not caused by flares. The predicted complete stripping of AU Mic d will place it among the population of stripped planets thought to make up most exoplanets with radii below the radius gap. Our results indicate that flares are likely to have a low impact on the location of this gap for the M dwarf planetary population since they do not significantly increase the mass loss rate of primordial atmospheres at the early stages of stellar and planetary evolution. 

If the planet were inside the HZ, the extra XUV emission from flares drives atmospheric depletion faster than the quiescent levels alone. This will occur especially during the post-saturation phase, which, in our simulations, is the evolutionary stage where the contribution of the flares reaches roughly 30$\%$ of the total XUV emission from the star. However, if the planet is located farther away from the star ($>$ 0.365 AU, receiving a flux of $\leq$ 0.8 W/m$^{2}$), not even the extra emission from stellar flares will drive the total atmospheric loss.

Our results also show that the only planet that has lost its primordial atmosphere is AU Mic d, which is not the closest planet but is the smallest in mass and radius. The lower gravity of the AU Mic d facilitates the process of atmospheric loss of the planet even when compared to AU Mic b, which is closer to the star but is also bigger in mass and size than AU Mic d. Although AU Mic b's escape rate is about 100 times higher than AU Mic d, its hydrogen envelope of 0.78 M$_{\oplus}$ prevents its full depletion. Similarly, AU Mic c not only has higher gravity and a larger hydrogen inventory but is also located further away from the host star compared to AU Mic b and d. As a result, even though AU Mic c's escape rate is about the same as AU Mic d, the atmosphere of AU Mic c is not significantly affected by photoevaporation.

Compared with the stellar quiescent emission at such a young age (i.e., $\sim$ 23 Myrs), the results considering flares do not show a significant difference in the atmospheric evolution of the planets. To understand what scenario maximizes the role of flares in atmospheric escape, we ran simulations moving AU Mic d to different distances from the host star, exposing the planet to different XUV fluxes. We found that for an early-type M star, such as AU Mic, flares will accelerate the escape for Earth-size planets in the HZ in a few thousand years. When placed halfway through the HZ (about 0.365 AU), the planet will completely lose its atmosphere only if flares are present. This result implies that close-in planets such as LHS 1815 b, K2-257 b, and TOI-4527.01 which are orbiting M1-2 V stars and have a size similar to AU Mic d, might have lost their atmospheres very early in their evolution. However, super-Earths orbiting M1-2 V stars farther from the HZ, such as Kepler-296 b and e, Kepler-186 b, Kepler-369 c, and Kepler-974 b, might lose their primordial atmospheres late in life due to stellar flares, if the flares of their hosts follow the same behavior of the XUV FFDs which we adopted in this work. Further investigation is necessary to observationally validate these FFDs and constrain the relevance of flares across different stellar ages, spectral types, and orbital planetary distances. Finally, telescopes such as the James Webb Space Telescope and the Giant Magellan Telescope should prioritize young terrestrial planets to observe their primordial atmospheres while they remain intact.

We very much thank the anonymous referee for the careful and thoughtful comments, which have certainly improved the paper. L.A. would like to thank Antígona Segura, Jacob Klemm, Jamie Dietrich, Kosuke Namekata, Rachel Osten, Rory Barnes, Ruth Murray-Clay, and Terry Christenson for useful insights. L.A. and E.S. acknowledge support from the NASA Virtual Planetary Laboratory Team through grant number 80NSSC18K0829. This material is based upon work performed as part of the CHAMPs (Consortium on Habitability and Atmospheres of M-dwarf Planets) team, supported by the National Aeronautics and Space Administration (NASA) under grant nos. 80NSSC21K0905 and 80NSSC23K1399 were issued through the Interdisciplinary Consortia for Astrobiology Research (ICAR) program. S.P. acknowledges support from NASA under award number 80GSFC24M0006.

\vspace{5mm}
\facilities{Exoplanet Archive \citep{exoplanetarchive}. Last Accessed: 2025-01-11;\\
The Planetary Habitability Laboratory @ UPR Arecibo (phl.upra.edu). Last Accessed: 2025-01-11.}

\software{NumPy \citep{harris2020array},
          Matplotlib \citep{Hunter2007},
          \vplanet \citep{barnes2020vplanet},
          \flare \citep{amaral2022},
          seaborn \citep{Waskom2021},
          pandas \citep{mckinney2010},
          SciPy \citep{2020SciPy}.}

\bibliography{references}{}
\bibliographystyle{aasjournal}

\end{document}